\documentclass[aps,prb,reprint,superscriptaddress,amsmath,amssymb,showpacs]{revtex4-1}%

\renewcommand{\vec}[1]{\mathbf{#1}}

\usepackage{braket}
\usepackage{subfigure}
\usepackage{graphicx}
\usepackage{xcolor}
\usepackage{dsfont}
\usepackage{pstricks}
\usepackage{mathrsfs}
\usepackage{bbm}
\usepackage{nicefrac}

\begin{document}

\title{Non-Markovian effects in the spin-boson model at zero temperature}

\author{S.\ Wenderoth}
\affiliation{Institute of Physics, University of Freiburg, Hermann-Herder-Str.\ 3, D-79104 Freiburg, Germany
}


\author{H.-P. Breuer}
\affiliation{Institute of Physics, University of Freiburg, Hermann-Herder-Str.\ 3, D-79104 Freiburg, Germany
}
\affiliation{EUCOR Centre for Quantum Science and Quantum Computing, University of Freiburg, Hermann-Herder-Str. 3, D-79104 Freiburg, Germany
}

\author{M.\ Thoss}
\affiliation{Institute of Physics, University of Freiburg, Hermann-Herder-Str.\ 3, D-79104 Freiburg, Germany
}
\affiliation{EUCOR Centre for Quantum Science and Quantum Computing, University of Freiburg, Hermann-Herder-Str. 3, D-79104 Freiburg, Germany
}

\date{\today}

\begin{abstract}
We investigate memory effects in the spin-boson model using a recently proposed measure for non-Markovian behavior based on the information exchange between an open system and its environment. Employing the numerical exact multilayer multiconfiguration time-dependent Hartree approach, we simulate the dynamics of the spin-boson model at zero temperature for a broad range of parameters. For a fast bath, i.e.\ in the scaling limit, we find non-Markovian dynamics for a coherently decaying spin at weak system-bath coupling, whereas memory effects are absent for stronger coupling in the regimes of incoherent decay and localization. If the time scales of system and bath are comparable, a complex, non-monotonic dependence of non-Markovianity on the system-bath coupling strength is observed.

\end{abstract}


\maketitle


Open quantum systems are characterized by exchange of particles, energy or information with an environment and are ubiquitous in physics and chemistry.\cite{Weiss1999,Breuer2007}
The coupling to the environment induces decoherence and dissipation, causing the relaxation of the system to an equilibrium or steady state. Besides these well-understood effects, the environment can also act as a memory for the open system leading to non-Markovian dynamics in the time evolution of the open system. 
Memory effects are apparent, for example, in the Nakajima-Zwanzig equation for the reduced density matrix of an open quantum system, which involves a convolution with a memory kernel.\cite{Nakajima1958,Zwanzig1960,Mori1965,Kidon2018}
A rigorous and representation-independent characterization and quantification of non-Markovianity 
in quantum systems is, however, challenging because concepts developed in classical probability theory cannot be applied.\cite{Breuer2016}
In order to investigate quantum non-Markovianity, different mathematical and physical concepts have been developed in recent years,\cite{Breuer2016} based on the divisibility of the dynamical map or correlations with an ancilla system.\cite{Wolf2008,Rivas2010,Chruscinski2011,Luo2012a} Here, we will employ a concept based on the flow of information between the open system and its environment.\cite{Breuer2009, Laine2010a} The central quantity in this approach is the trace distance between two quantum states of the open system,\cite{Nielsen2000,Hayashi2006}
\begin{align}
\mathcal{D}(\rho_1, \rho_2) = \frac{1}{2}{\rm Tr}|\rho_1-\rho_2|, \label{eq:trace_distance}
\end{align} 
where the modulus of an operator $A$ is given by $|A|=\sqrt{A^\dagger A}$. This quantity can be interpreted as the distinguishability between the two states $\rho_1$ and $\rho_2$.\cite{Fuchs1999} Assuming that the initial state factorizes between the open system and the environment, the time evolution of the open system is determined by a family of completely positive and trace preserving (CPT) maps $\Phi(t)$. Any pair of initial states $\rho_{1,2}(0)$ then evolves into $\rho_{1,2}(t)=\Phi(t)\rho_{1,2}(0)$. The time-dependent trace distance is defined as
\begin{align}
\mathcal{D}(t) = \mathcal{D}(\rho_1(t), \rho_2(t)).\label{eq:time_tdp_trace-dist}
\end{align}
Note that CPT maps are contractions for the trace distance, i.e. $\mathcal{D}(t) \leq \mathcal{D}(0)$.\cite{Ruskai1994} The CPT property alone, however, does not imply monotonicity of the trace distance as a function of time. If $\mathcal{D}(t)$ is a monotonically decreasing function of time and, hence, the two states $\rho_1(t)$ and $\rho_2(t)$ become less and less distinguishable, which can be interpreted as a continuous loss of information from the system to the environment, the dynamics is defined to be Markovian. Correspondingly, a temporal increase of the trace distance can be interpreted as a flow of information from the environment back to the open system, which is a unique signature of memory effects and, thus, of the non-Markovian character of the dynamics. On the basis of this interpretation one can define a measure for the degree of non-Markovianity of the dynamics by means of\cite{Breuer2009}
\begin{align}
\mathcal{N} &= \max_{\rho_{1,2}(0)}\int_{\sigma>0} {\rm d}t ~\sigma(t),\label{eq:summed_non_Markovianity}
\end{align}
where $\sigma(t)=\frac{d}{dt}\mathcal{D}(t)$ and the integral extends over all time intervals in which $\sigma(t)>0$. By definition, this measure is strictly zero if the trace distance decreases monotonically, i.e.\ if there is no information backflow from the environment to the system, which corresponds to Markovian dynamics. Such a behaviour occurs, e.g., if the family of dynamical maps $\Phi(t)$ is CP-divisible.\cite{Breuer2016} The simplest and best known example is given by a dynamical semigroup with a time-independent generator in Lindblad form.

In this paper, we employ the above discussed measure for non-Markovianity to investigate memory effects in the spin-boson model at zero temperature. 
To the best of our knowledge, this is the first systematic, non-perturbative (numerically exact) study of 
non-Markovianity in a non-integrable model in the deep quantum regime at zero temperature. Previous studies of non-Markovianity in the spin-boson model used perturbative approaches or focused on special parameter regimes such as higher temperature.\cite{Clos2012,Chen15,Rivas17,Hinarejos17,Kurt18}  
We also mention a path-integral study of non-Markovianity in a related model,\cite{Thorwart2013} albeit for finite temperature, and a very recent, similar investigation in the integrable model of quantum Brownian motion.\cite{Einsiedler20}

The spin-boson model, which involves a two-level system (or spin) interacting linearly with a bath of harmonic oscillators, is a paradigmatic model to describe dissipative quantum dynamics.\cite{Leggett1987,Weiss1999} Despite its simple form, it has applications to a variety of different processes and phenomena, including electron transfer\cite{Marcus1985} and macroscopic quantum coherence\cite{Weiss1987}.
On the other hand, the spin-boson model is also interesting from a more fundamental point of view as it shows a transition from coherent dynamics to incoherent decay as well as a quantum phase transition.\cite{Bray1982,Chakravarty1982,Wang2019} Here, we focus on the unbiased spin-boson model. Employing mass-weighted coordinates, the Hamiltonian reads
\begin{align}
H &= \Delta\sigma_x + \frac{1}{2}\sum_n (p_n^2 + \omega_n^2 q_n^2) + \sigma_z \sum_n c_n q_n,
\end{align} 
where $\sigma_x$ and $\sigma_z$ are the Pauli matrices, $\Delta$ denotes the coupling between the two spin states, and $\omega_n$, $q_n$, and $p_n$ represent the frequency, position and momentum of the bath oscillators, respectively. The properties of the bath which influence the spin are summarized by the spectral density\cite{Leggett1987,Weiss1999}
\begin{align}
J(\omega)&=\frac{\pi}{2}\sum_n \frac{c_n^2}{\omega_n}\delta(\omega-\omega_n).
\end{align}
Here, we consider a spectral density of Ohmic form with an exponential cutoff
\begin{align}
J(\omega)&= \frac{\pi}{2}\alpha \omega{\rm e}^{-\nicefrac{\omega}{\omega_c}},
\end{align}
where $\alpha$ defines the coupling strength and $\omega_c$ denotes the characteristic frequency of the bath. In the scaling limit ($\omega_c /\Delta \to \infty$), the dynamics of the spin can be grouped into three qualitatively different regimes, comprising coherent decay for weak system-environment coupling ($\alpha<0.5$), incoherent decay (intermediate coupling, $0.5<\alpha<1$) and localization (strong coupling $\alpha >1$).  It is also known that for finite
$\omega_c /\Delta$
both critical couplings $\alpha$ shifts to larger values.\cite{Thoss2001,Wang2008,Wang10,Wang2019}


To simulate the dynamics of the spin-boson model, we use the multilayer multiconfiguration time-dependent Hartree approach (ML-MCTDH)\cite{Wang2003,Manthe08,Vendrell11,Wang2015} which allows to propagate the wave function of the joint system in a numerically exact way.
The ML-MCTDH approach represents a rigorous variational basis-set method, which uses a multiconfiguration expansion of the wave function, employing time-dependent basis functions and a hierarchical multilayer representation. Within this framework the wave function is expanded in terms of time-dependent configurations,
\begin{align}
	\ket{\Psi(t)} = \sum_{\mathbf{J}} A_{\mathbf{J}}(t) \prod_{n=1}^N \ket{\phi^n_{j_n}(t)},
\end{align}
where $\mathbf{J}$ is a $N$-dimensional multi-index. Each configuration is a Hartree product of 'single-particle' functions (SPFs) $\ket{\phi^n_{j_n}(t)}$, where $N$ denotes the total number of single-particle (SP) degrees of freedom and $n$ is the index of a particular SP group. Each SPF is again expanded in terms of time-dependent configurations similar to the original wave function,
\begin{align}
	&\ket{\phi^n_{j}(t)} = \sum_{\mathbf{I}} B_{\mathbf{I}}^{n,j}(t) \prod_{q=1}^{Q(n)} \ket{\nu^{(n,q)}_{i_q}(t)},
\end{align}
where $Q(n)$ denotes the number of level two (L2) SP groups in the $n$th level one (L1) SP group and $\ket{\nu^{(n,q)}_{i_q}(t)}$ is the L2-SPF for the $q$th L2-SP group. Similarly, the L2-SPFs can be expanded in a multiconfigurational expansion. As a result, the overall wave function $\ket{\Psi(t)}$ is expanded recursively to many layers in the ML-MCTDH framework, a representation which corresponds to a hierarchical tensor decomposition in the form of a tensor tree network. Following the Dirac-Frenkel variational principle,\cite{Frenke1934} the equations of motion are obtained from a variation of the wave function $\ket{\Psi(t)}$ with respect to the expansion coefficients of each layer.\cite{Wang2009} The ML-MCTDH approach allows for the simulation of large but finite quantum systems. Thus, we represent the continuous bath by a finite number of modes. In this work, we use an equidistant distribution but other choices are possible.\cite{Wang2008,Vega2015} To ensure convergence to the continuum limit over the time scale considered, we employ several hundred of modes. For a detailed discussion of the numerical treatment of a continuous bath see Ref.\ \onlinecite{Wang2008}.

Using the ML-MCTDH approach, we investigate non-Markovian behaviour in the spin-boson model. We focus on the low temperature regime, where non-Markovian effects are expected to be particularly pronounced.\cite{Clos2012} To evaluate the time-dependent trace distance defined in Eq.\:(\ref{eq:time_tdp_trace-dist}), we simulate the dynamics for two different initial states of the spin. The reduced state of the spin is uniquely described by the Bloch vector $\vec{a}(t) = (\braket{\sigma_x}(t), \braket{\sigma_y}(t), \braket{\sigma_z}(t))^T$, where $\braket{\sigma_i}(t) = \braket{\psi(t)|\sigma_i |\psi(t)}$. Employing this representation, Eq.\:(\ref{eq:time_tdp_trace-dist}) can be expressed as
\begin{align}
\mathcal{D}(t) = \frac{1}{2}|\vec{a}_1(t) - \vec{a}_2(t)|,
\end{align}
where $\vec{a}_{1(2)}(t)$ is the time-dependent Bloch vector corresponding to the first (second) initial state, respectively, and  $|\vec{a}_1 - \vec{a}_2|$ denotes the Euclidean distance. It can be shown\cite{Wissmann2012} that initial system states leading to a maximal non-Markovianity $\mathcal{N}$ must have the maximal initial trace distance $\mathcal{D}(0)=1$ and, hence, must have orthogonal supports. For the present case of a two-state system (spin) this implies that the initial states must be a pair of pure orthogonal states (antipodal points on the Bloch sphere). To avoid the numerically expensive maximization over all possible initial states, we fix the initial states to the two eigenstates of $\sigma_z$, $\rho_1(0)=\ket{\uparrow}\bra{\uparrow}$ and $\rho_2(0)=\ket{\downarrow}\bra{\downarrow}$. In previous studies this choice gave a reasonable lower bound for the non-Markovianity.\cite{Clos2012} The harmonic oscillators are initially all in the ground state and there is no correlation between the spin and the environment.

For the chosen initial states, the time evolution of $\vec{a}_2$ can be related to that of $\vec{a}_1$ (see supplementary material). Using this and the relation $\braket{\sigma_y}(t)=\frac{1}{2\Delta}\partial_t\braket{\sigma_z}(t)$, the trace distance can be written in terms of $\braket{\sigma_z}_1$ and its derivative as
\begin{align}
\mathcal{D}(t) &= \sqrt{\big[\braket{\sigma_z}_1(t)\big]^2 + \big[\frac{1}{2\Delta}\partial_t \braket{\sigma_z}_1(t)\big]^2}. \label{eq:general_trace_distance}
\end{align}
As a consequence of this equation, it follows that $\partial_t \braket{\sigma_z}_1(t_{\rm s})=0$ implies $\partial_t \mathcal{D}(t_{\rm s})=0$, i.e.\ if $\braket{\sigma_z}_1$ has a stationary point at $t_{\rm_s}$, the trace distance also has a stationary point at $t_{\rm_s}$. 

For later analysis, we note that in the weak coupling and large $\omega_c$ limit, an approximate analytic solution\cite{Weiss1999} for $\braket{\sigma_z}$ can be used to derive the following equation for the trace distance,
\begin{align}
\mathcal{D}(t) &= {\rm e}^{-\gamma t} \sqrt{\frac{1}{2} \big[1+\eta\big] + \beta \sin(2\tilde{\Delta} t) +\frac{1}{2} \big[1-\eta\big] \cos(2\tilde{\Delta} t)}.\label{eq:weak_coupling_trace_distance}
\end{align}
Here, $\gamma$, $\tilde{\Delta}$, $\beta$, and $\eta$ are constants which depend on the coupling strength $\alpha$ and on the characteristic bath frequency $\omega_c$. The explicit expressions are given in the supplementary material. 
		
\begin{figure}[h]
\hspace*{-0.5cm}
\includegraphics[scale=0.5]{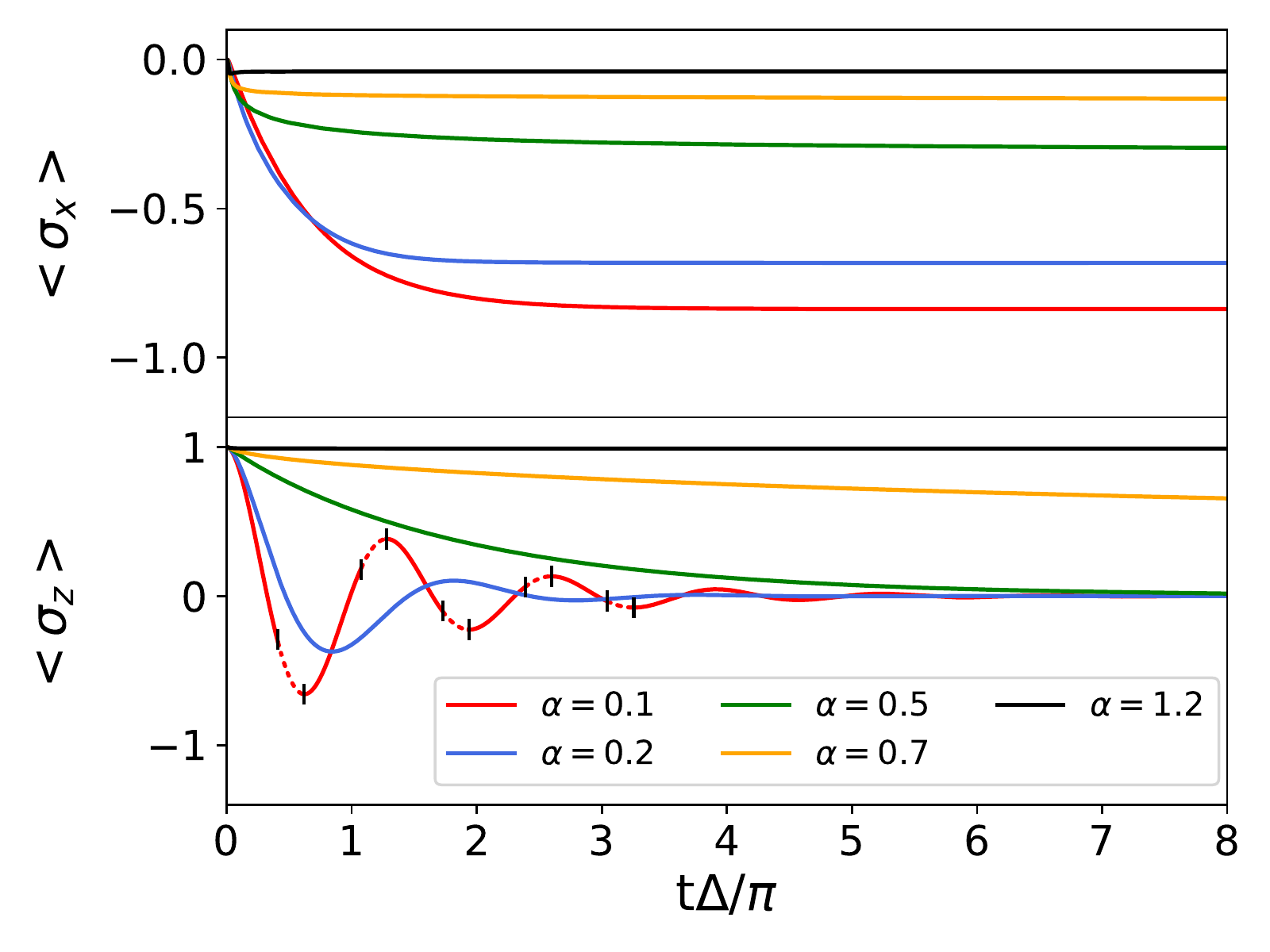}
\caption{Expectation value of $\sigma_x$ and $\sigma_z$ in the three different dynamical regimes of the spin-boson model for $\omega_c=40\Delta$. For $\alpha=0.1$ the first five non-Markovian intervals are marked by dashed lines.}\label{fig:spin_dynamics}
\end{figure}

We begin our discussion of non-Markovian effects in the scaling regime, i.e.\ $\omega_c\gg \Delta$. As a starting point, we recapitulate the dynamics of the spin and discuss the corresponding dynamics of the trace distance.\cite{Leggett1987,Weiss1999,Wang2008} Fig.\ \ref{fig:spin_dynamics} shows the dynamics of the spin for different values of the coupling strength $\alpha$ for $\omega_c=40\Delta$, demonstrating the three qualitatively different dynamical regimes. Since $\braket{\sigma_y}(t)=\frac{1}{2\Delta}\partial_t\braket{\sigma_z}(t)$ holds, we only present the dynamics of $\braket{\sigma_x}$ and $\braket{\sigma_z}$.
In the weak coupling regime ($\alpha<0.5$) the spin decays coherently to its stationary value. For increasing coupling strength, the oscillation frequency of $\braket{\sigma_z}$ decreases. The intermediate regime ($0.5 < \alpha<1$) is characterized by a monotonic or incoherent decay of the spin. Upon increasing the coupling strength further, the decay slows down and eventually the spin localizes for coupling strengths larger than  $\alpha=1$. In all three regimes $\braket{\sigma_x}$ relaxes monotonously  to its equilibrium value. For a more comprehensive discussion of the dynamics of the spin-boson model at zero temperature, see e.g.\ Refs.\ \onlinecite{Wang2008,Wang10}.

\begin{figure}[h]
\hspace*{-0.5cm}
\includegraphics[scale=0.5]{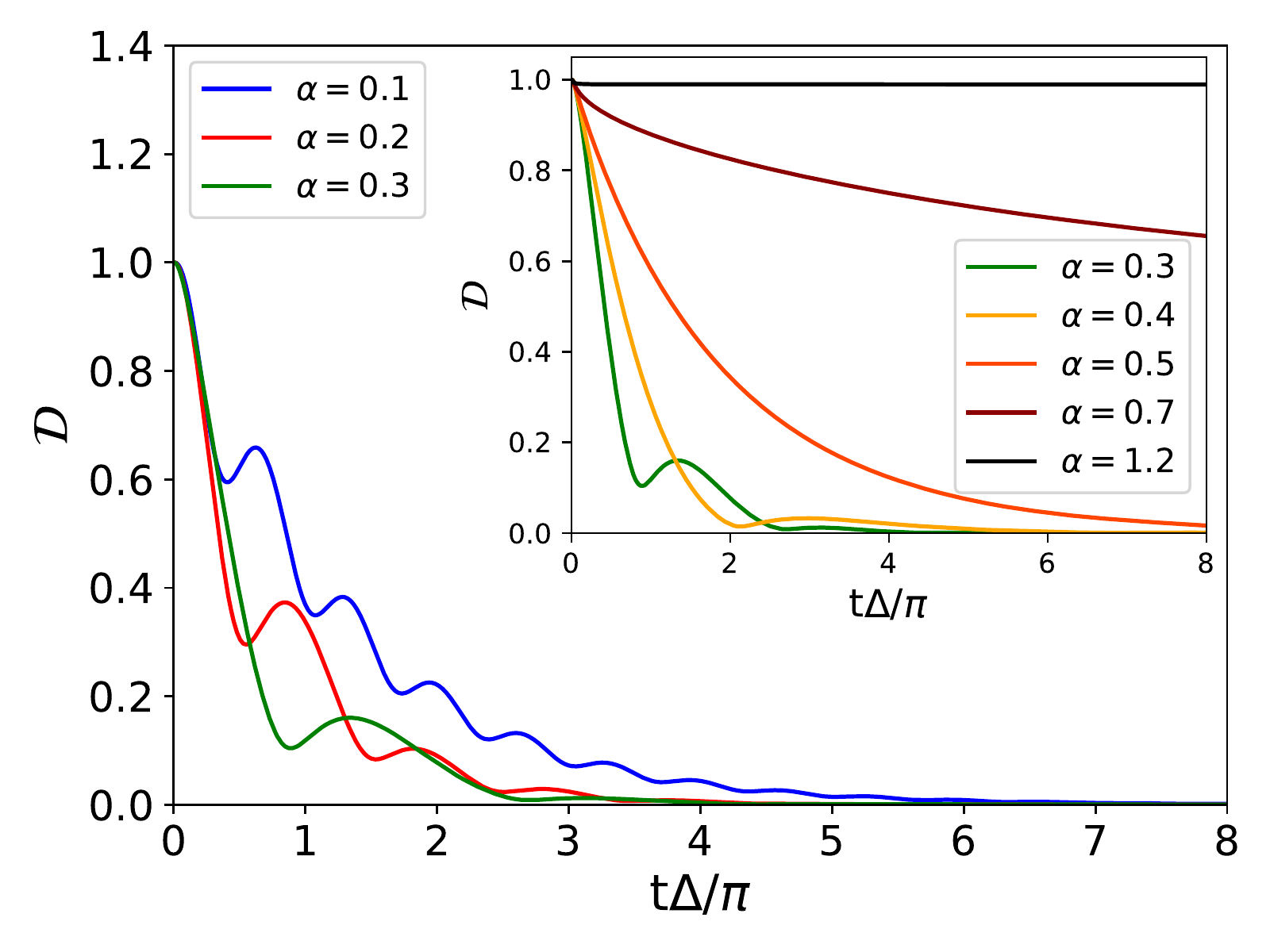}
\caption{Dynamics of the trace distance for different values of $\alpha$ and $\omega_c=40\Delta$.}\label{fig:trace_dynamics}
\end{figure}

Next, we analyze the trace distance, which is the central object to quantify memory effects. Fig.\ \ref{fig:trace_dynamics} shows the time-dependent trace distance for different values of the coupling strength $\alpha$ and $\omega_c=40\Delta$. The behavior of the trace distance can be grouped into three different regimes, similar to the dynamics of the spin itself.
For weak coupling ($\alpha\lesssim0.5$), the trace distance exhibits an overall decay to zero with periodic modulations including temporal increases, which indicate the presence of memory effects. For intermediate coupling in the incoherent regime of spin dynamics ($0.5\lesssim\alpha\lesssim1$), the trace distance decays monotonically. The decay slows down upon increasing coupling strength. In both regimes, the overall decay reflects a relaxation of the spin to an equilibrium state, which is independent on the initial state of the spin. In the strong coupling, localized regime ($\alpha>1$) the spin is frozen in its initial state, and thus, the trace distance remains close to its initial value of one. We conclude that in the scaling regime the spin exhibits non-Markovian dynamics for $\alpha\lesssim0.5$. In the following, we discuss this non-Markovian regime in more detail.

For a coherently decaying spin, $\braket{\sigma_z}(t)$ exhibits local minima and maxima and, thus, the trace distance $\mathcal{D}(t)$ has stationary points. Employing Eq.\ (\ref{eq:weak_coupling_trace_distance}), it can be shown that these stationary points are all local maxima and, therefore, the non-Markovian intervals end at the extremal points of $\braket{\sigma_z}$. This is demonstrated in Fig.\ \ref{fig:spin_dynamics} for $\alpha=0.1$. We find this behaviour for all couplings $\alpha<0.5$, as long as $\omega_c\gg\Delta$. As the coupling strength approaches the coherent-to-incoherent transition ($\alpha=0.5$), the renormalized frequency $\tilde{\Delta}$ vanishes, leading to a monotonically decaying spin. Additionally, the increases in the trace distance become weaker as $\alpha\to 0.5$. Thus, the non-Markovian intervals shift to infinite time and memory effects disappear as the dynamics changes from coherent to incoherent decay. As a result, the dynamics is Markovian for $\alpha\geq0.5$.

\begin{figure}
\hspace*{-0.5cm}
\includegraphics[scale=0.5]{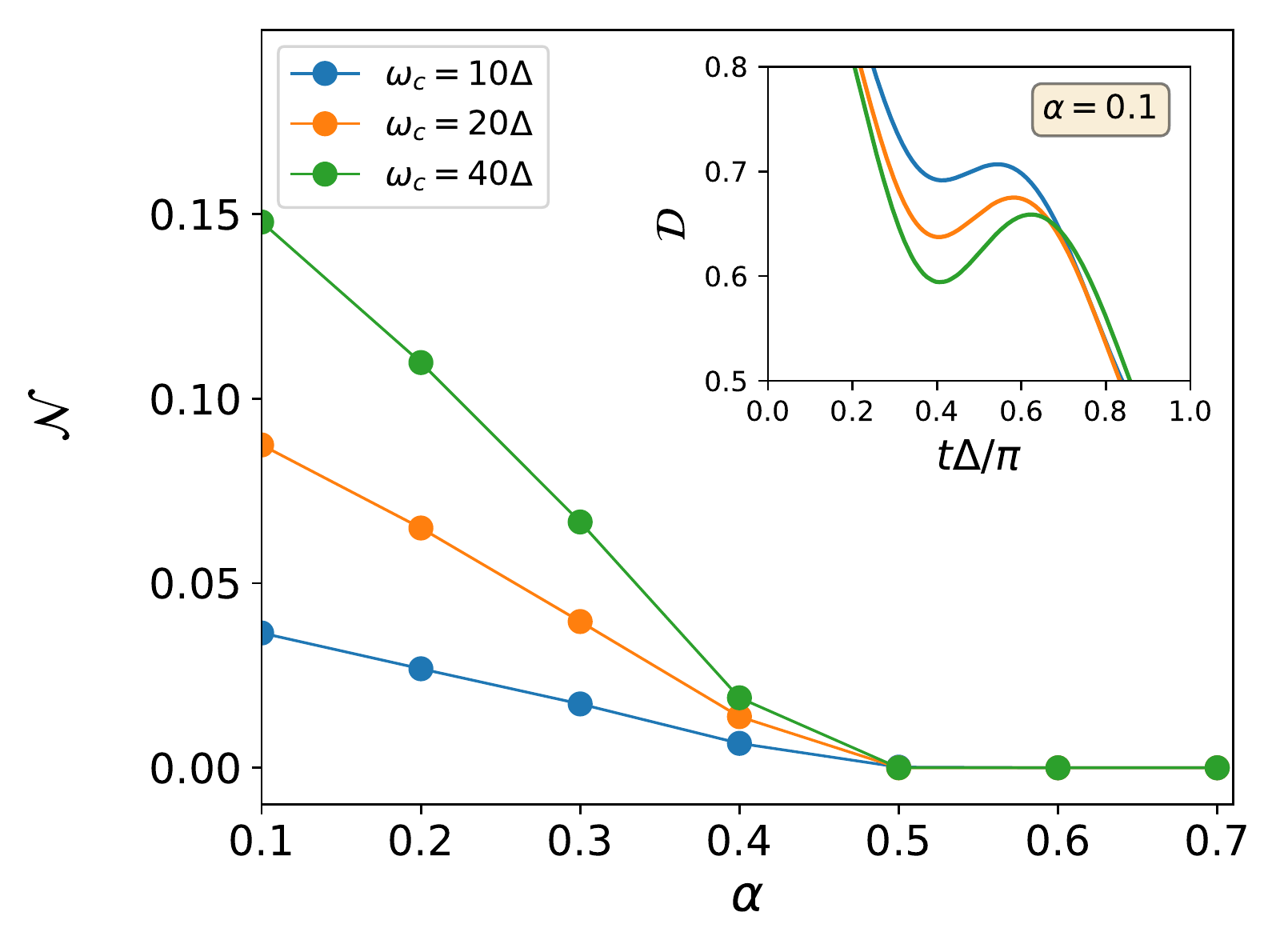}
\caption{Non-Markovianity $\mathcal{N}$ as a function of the coupling strength $\alpha$ for different characteristic frequencies $\omega_c$. The symbols represent the data, the lines are a guide to the eye. The inset shows the first increase of the trace distance for different values of $\omega_c$ for $\alpha=0.1$.}\label{fig:summed_non_markovianity}
\end{figure}

In order to quantify the non-Markovianity as a function of the coupling strength $\alpha$, we use the cumulative measure $\mathcal{N}$ defined in Eq.\:(\ref{eq:summed_non_Markovianity}). First note that without system-bath coupling (i.e.\ $\alpha=0$), $\mathcal{D}(t)=1$ holds for all times and, thus, $\mathcal{N}=0$, as expected for a unitary time-evolution. Fig.\ \ref{fig:summed_non_markovianity} shows $\mathcal{N}$ as a function of the coupling strength $\alpha$ for different values of $\omega_c$. 
Because the trace distance exhibits memory effects in the regime of coherent decay, i.e. for $\alpha<0.5$, the measure of non-Markovianity $\mathcal{N}$ is non-zero. In this regime, the cumulative information back flow decreases monotonically as a function of the coupling strength. For $\alpha<0.1$, the decay is too slow to obtain $\mathcal{N}$ directly from numerical simulations. Employing Eq.\ (\ref{eq:weak_coupling_trace_distance}), which is valid in this weak coupling regime, we find $\lim\limits_{\alpha \to 0} \mathcal{N} = \mathcal{N}_0>0$ for fixed $\omega_c$ and, thus, the non-Markovianity is not analytic at $\alpha=0$. The detailed derivation is provided in the supplementary material, which also provides a discussion of the validity of perturbative methods such as the time-convolutionless master equation\cite{Clos2012} to describe non-Markovian effects in the weak coupling regime.
For $\alpha\geq0.5$, the dynamics are Markovian and, consequently, the non-Markovianity $\mathcal{N}$ vanishes. 

We finish the discussion of the dynamics in the scaling regime with the influence of the time scale of the bath, determined by the characteristic frequency $\omega_c$, on the non-Markovian behavior of the spin, illustrated in Fig.\ \ref{fig:summed_non_markovianity}. For fixed $\alpha$, the memory effects are more pronounced for larger characteristic bath frequencies as can be seen in the inset of Fig.\ \ref{fig:summed_non_markovianity}. Consequently, the non-Markovianity $\mathcal{N}$ increases upon increasing the characteristic bath frequency. For all $\omega_c\geq 5\Delta$, we find a similar behavior of $\mathcal{N}$, i.e. $\mathcal{N}$ is non-zero only for $\alpha \leq 0.5$. Note that for $\omega_c=5\Delta$ the non-Markovianity is almost zero ($\mathcal{N}<0.01$) for all couplings $\alpha$ (data not shown).

In the following, we discuss non-Markovian effects outside the scaling limit, focusing on the particularly interesting crossover regime $\omega_c\approx\Delta$, where the timescales of spin and bath are similar.
\begin{figure}
\hspace*{-0.5cm}
\includegraphics[scale=0.48]{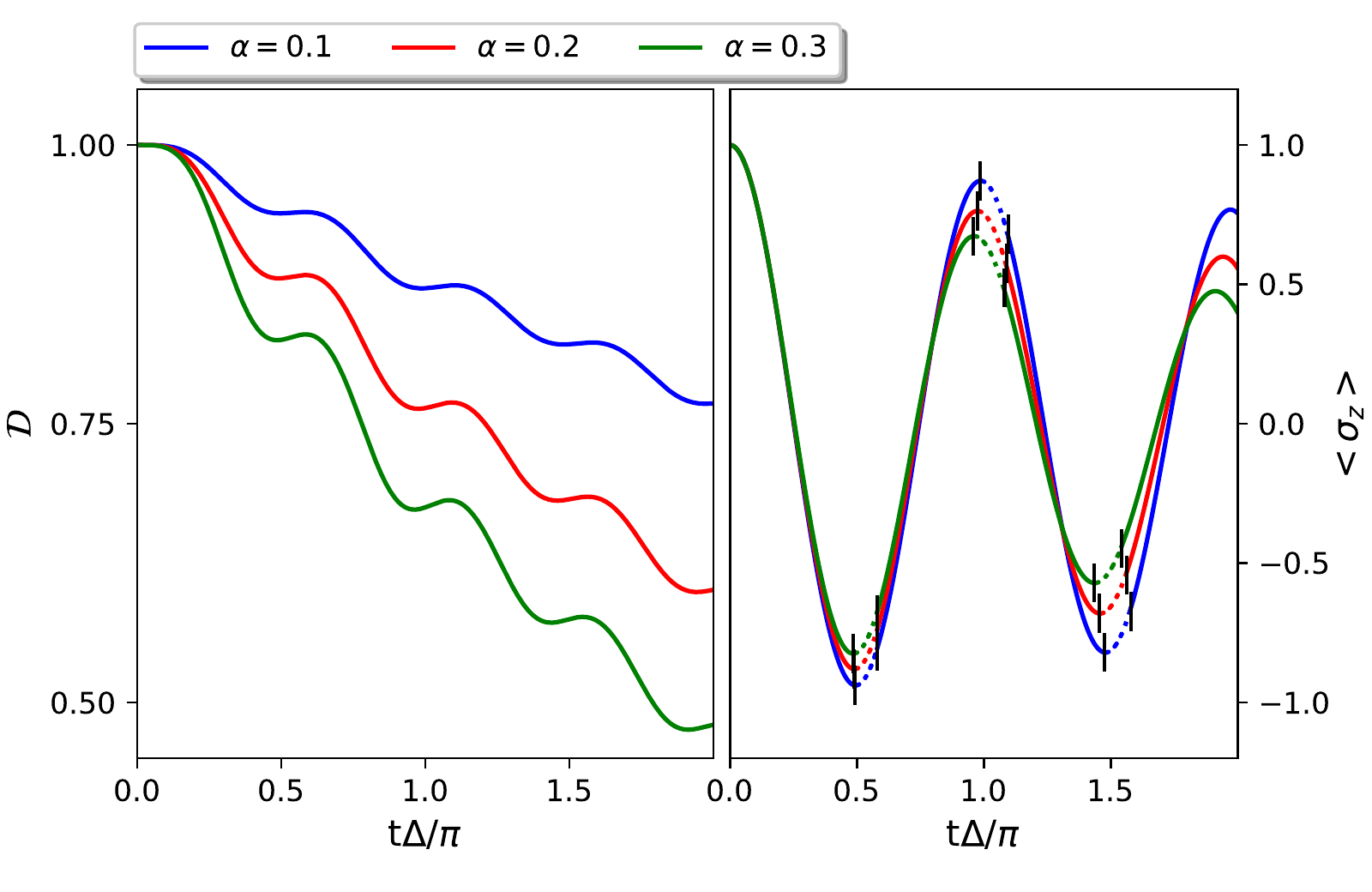}
\caption{Dynamics of the trace distance (left) and $\sigma_z$ (right) for $\omega_c=\Delta$ in the weak coupling regime. 
For better visualization, only results for short times are shown. The non-Markovian intervals of $\sigma_z$
are marked by dashed lines.}\label{fig:non-scaling_weak}
\end{figure}
For weak coupling $\alpha$, depicted in Fig.\ \ref{fig:non-scaling_weak} for the case $\omega_c=\Delta$,
the overall dynamics
are very similar to the scaling regime, i.e. the spin shows damped, coherent oscillations and the trace distance decays to zero with periodic modulations.
The memory effects, however, exhibit a qualitatively different behavior. Unlike in the scaling regime, 
the non-Markovian intervals begin at the extremal points of $\braket{\sigma_z}$, indicating that $\mathcal{D}$ has local minima at the extremal points of $\braket{\sigma_z}$.

\begin{figure}
\hspace*{-0.5cm}
\includegraphics[scale=0.48]{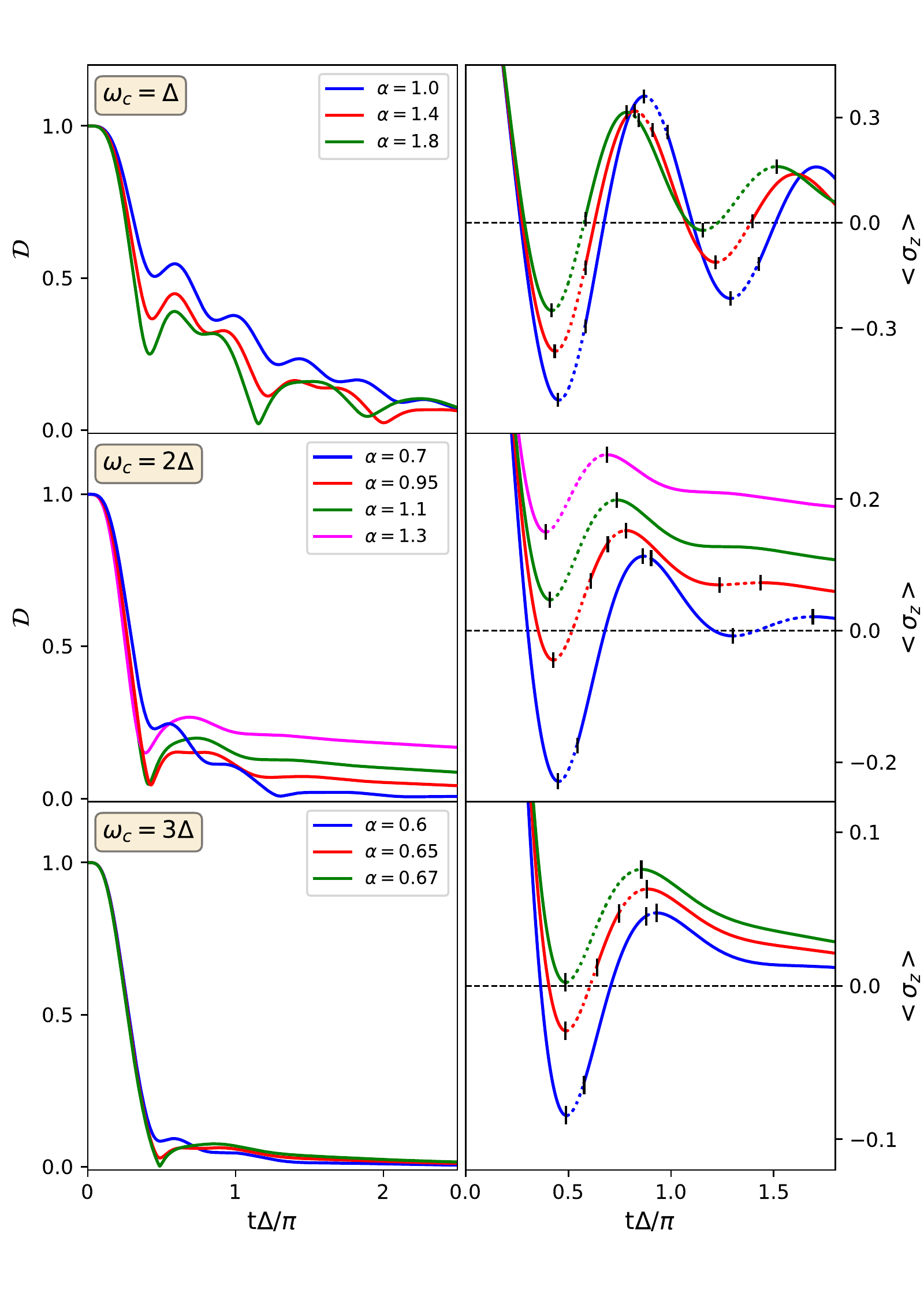}
\caption{Dynamics of the trace distance (left) and $\sigma_z$ (right) for different characteristic frequencies $\omega_c$ in the moderate and strong coupling regime. 
The non-Markovian intervals are marked with dashed lines. For better visualization not all non-Markovian intervals are marked.}\label{fig:non-scaling_strong}
\end{figure}

Further differences to the scaling regime are observed for stronger coupling $\alpha$ depicted in Fig.\ \ref{fig:non-scaling_strong}. The dynamics in this regime depends sensitively on $\omega_c$ and, therefore, we show results for different values of $\omega_c$.
Different to the scaling regime, the dynamics of the spin is partially coherent for $\alpha\geq0.5$ (see also Refs.\ \onlinecite{Thoss2001,Wang2008}), with an initial decay which does not slow down as the coupling strength is increased. 

In addition, we find a qualitatively different non-Markovian behavior in the crossover regime, $\omega_c\approx\Delta$. To demonstrate this, consider the first local minimum and maximum of $\braket{\sigma_z}$ for the case $\omega_c=2\Delta$ in Fig.\ \ref{fig:non-scaling_strong} (similar for $\omega_c=3\Delta$). In the weak  to moderate coupling regime (up to $\alpha\approx 0.7$), the stationary points of $\mathcal{D}$ associated to the two local extrema of $\braket{\sigma_z}$ are both local minima and, thus, both non-Markovian intervals begin at the local extrema. As the coupling strength is increased, the non-Markovian interval associated to the local maximum of $\braket{\sigma_z}$ first shrinks to zero and then extends to the left (i.e.\ to shorter times) with fixed end point at the local maximum of $\braket{\sigma_z}$. Eventually the two distinct intervals merge to a single non-Markovian interval extending from the local minimum to the local maximum. Upon further increasing the coupling strength, the initial decay becomes weaker and the following increase in the trace distance becomes smaller. This transition in the non-Markovian behavior results in a non-monotonic dependence of the memory effects on the coupling strength $\alpha$ caused by this pair of extrema of $\braket{\sigma_z}$, i.e.\ memory effects first increase with $\alpha$ up to the point at which the two intervals merge and then decrease again. Similar transitions in the non-Markovian behavior are observed for later local extrema of $\braket{\sigma_z}$, albeit for weaker coupling $\alpha$.  For smaller $\omega_c$, represented in Fig.\ \ref{fig:non-scaling_strong} by the case $\omega_c=\Delta$, the first two non-Markovian intervals do not show such a transition and, therefore, memory effects increase monotonically with the coupling strength.

\begin{figure}
\includegraphics[scale=0.48]{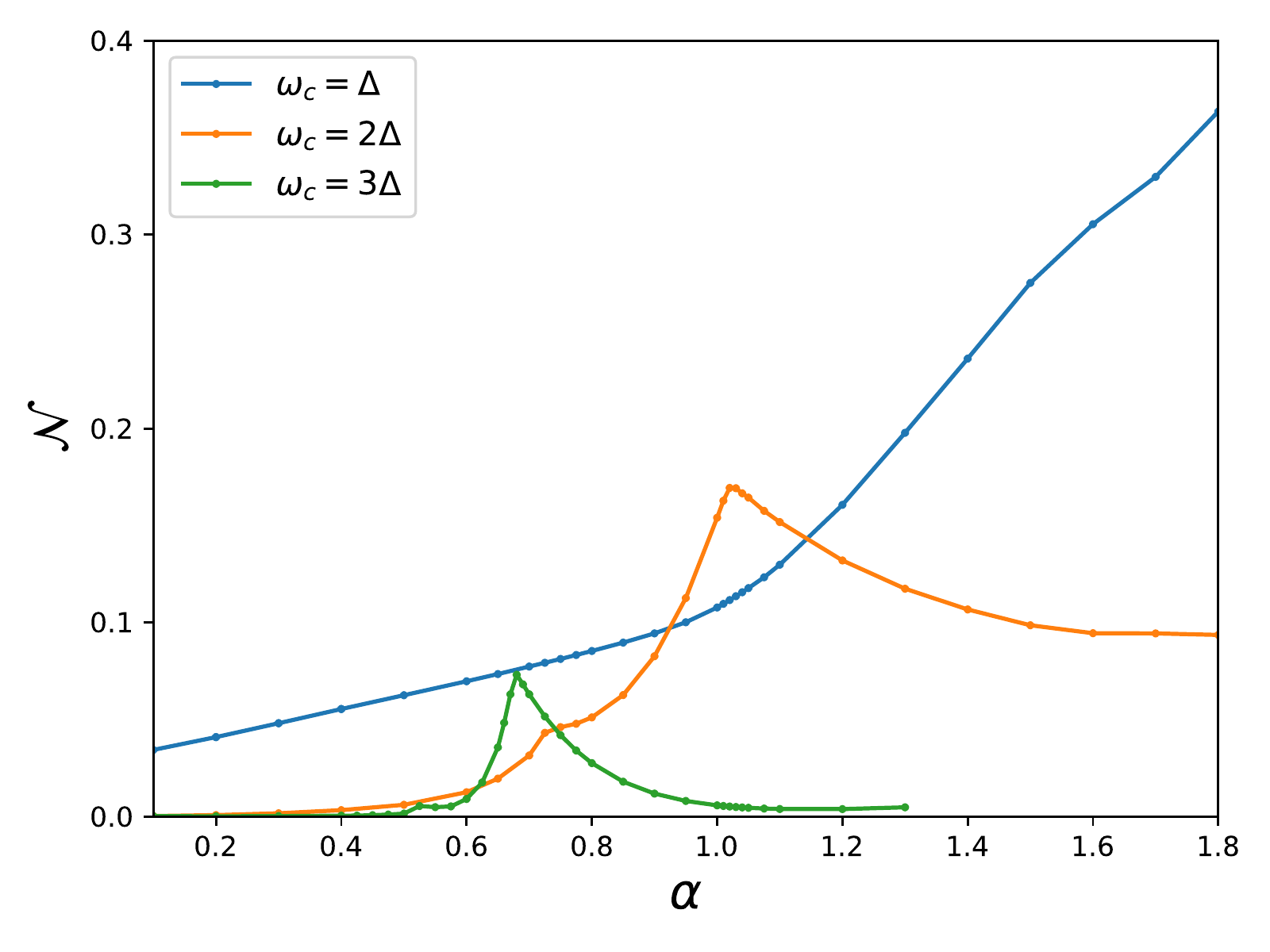}
\caption{Non-Markovianity $\mathcal{N}$ as a function of the coupling strength $\alpha$ for different characteristic frequencies $\omega_c$.}\label{fig:slow_bath_summed_nmm}
\end{figure}

This dependence of memory effects on the coupling strength $\alpha$ is reflected in the cumulative measure for non-Markovianity  $\mathcal{N}$ 
depicted in Fig.\ \ref{fig:slow_bath_summed_nmm}.
For weak coupling $\alpha$, $\mathcal{N}$ is very small for $\omega_c \in [2\Delta, 5\Delta]$, whereas it assumes significant values for $\omega_c=\Delta$. This is consistent with previous investigations of memory effects in the spin-boson model employing perturbative master equations, which predict Markovian dynamics if the effective spectral density of the environment is flat around the transition frequency of the spin,\cite{Clos2012} i.e.\ for $\omega_c=2\Delta$. In the moderate and strong coupling regime, the transitions in the non-Markovian behavior discussed above give rise to different features in the non-Markovianity measure. For $\omega_c=2\Delta$ and $\omega_c=3\Delta$, the transition of the first local minimum and maximum of $\braket{\sigma_z}$ lead to a pronounced maximum of $\mathcal{N}$. Additionally, we find structures at $\alpha\approx0.71$ and $\alpha\approx0.52$ for $\omega_c=2\Delta$ and $\omega_c=3\Delta$, respectively, which coincide with the transition of non-Markovian behavior of the second local minimum and maximum of $\braket{\sigma_z}$. For $\omega_c=\Delta$,  only the third oscillation of $\braket{\sigma_z}$ exhibits a transition at $\alpha\approx1.4$, leading to weak shoulder in the non-Markovianity $\mathcal{N}$. Otherwise $\mathcal{N}$ increases monotonically over the shown range of coupling strengths.


In summary, we have employed the numerically exact ML-MCTDH approach to investigate non-Markovian effects in the spin-boson model at zero temperature. The results obtained for a broad range of parameters reveal a rich dynamical behavior. While in the scaling limit of a fast bath, non-Markovian effects are limited to weak system-bath coupling, the crossover regime without separation of timescales between spin and bath exhibits a complex, non-monotonic dependence of non-Markovianity on the coupling strength. 
The question how these findings can be generalized to more complex systems and an interacting bath is an interesting subject for future work.

We thank Haobin Wang for providing the ML-MCTDH code used in this work and for many insightful discussions on quantum dynamics.
This work was supported by the German Research Foundation (DFG) through IRTG 2079 and FOR 5099. Furthermore, support by the state of Baden-{W\"urttemberg} through bwHPC and the DFG through grant no INST 40/575-1 FUGG (JUSTUS 2 cluster) is gratefully acknowledged.


\begin{thebibliography}{41}%
\makeatletter
\providecommand \@ifxundefined [1]{%
 \@ifx{#1\undefined}
}%
\providecommand \@ifnum [1]{%
 \ifnum #1\expandafter \@firstoftwo
 \else \expandafter \@secondoftwo
 \fi
}%
\providecommand \@ifx [1]{%
 \ifx #1\expandafter \@firstoftwo
 \else \expandafter \@secondoftwo
 \fi
}%
\providecommand \natexlab [1]{#1}%
\providecommand \enquote  [1]{``#1''}%
\providecommand \bibnamefont  [1]{#1}%
\providecommand \bibfnamefont [1]{#1}%
\providecommand \citenamefont [1]{#1}%
\providecommand \href@noop [0]{\@secondoftwo}%
\providecommand \href [0]{\begingroup \@sanitize@url \@href}%
\providecommand \@href[1]{\@@startlink{#1}\@@href}%
\providecommand \@@href[1]{\endgroup#1\@@endlink}%
\providecommand \@sanitize@url [0]{\catcode `\\12\catcode `\$12\catcode
  `\&12\catcode `\#12\catcode `\^12\catcode `\_12\catcode `\%12\relax}%
\providecommand \@@startlink[1]{}%
\providecommand \@@endlink[0]{}%
\providecommand \url  [0]{\begingroup\@sanitize@url \@url }%
\providecommand \@url [1]{\endgroup\@href {#1}{\urlprefix }}%
\providecommand \urlprefix  [0]{URL }%
\providecommand \Eprint [0]{\href }%
\providecommand \doibase [0]{http://dx.doi.org/}%
\providecommand \selectlanguage [0]{\@gobble}%
\providecommand \bibinfo  [0]{\@secondoftwo}%
\providecommand \bibfield  [0]{\@secondoftwo}%
\providecommand \translation [1]{[#1]}%
\providecommand \BibitemOpen [0]{}%
\providecommand \bibitemStop [0]{}%
\providecommand \bibitemNoStop [0]{.\EOS\space}%
\providecommand \EOS [0]{\spacefactor3000\relax}%
\providecommand \BibitemShut  [1]{\csname bibitem#1\endcsname}%
\let\auto@bib@innerbib\@empty
\bibitem [{\citenamefont {Weiss}(1999)}]{Weiss1999}%
  \BibitemOpen
  \bibfield  {author} {\bibinfo {author} {\bibfnamefont {U.}~\bibnamefont
  {Weiss}},\ }\href@noop {} {\emph {\bibinfo {title} {{Quantum Dissipative
  Systems}}}},\ \bibinfo {edition} {2nd}\ ed.\ (\bibinfo  {publisher} {World
  Scientific, Singapore},\ \bibinfo {year} {1999})\BibitemShut {NoStop}%
\bibitem [{\citenamefont {Breuer}\ and\ \citenamefont
  {Petruccione}(2007)}]{Breuer2007}%
  \BibitemOpen
  \bibfield  {author} {\bibinfo {author} {\bibfnamefont {H.-P.}\ \bibnamefont
  {Breuer}}\ and\ \bibinfo {author} {\bibfnamefont {F.}~\bibnamefont
  {Petruccione}},\ }\href@noop {} {\emph {\bibinfo {title} {{The Theory of Open
  Quantum Systems}}}}\ (\bibinfo  {publisher} {Oxford University Press,
  Oxford},\ \bibinfo {year} {2007})\BibitemShut {NoStop}%
\bibitem [{\citenamefont {Nakajima}(1958)}]{Nakajima1958}%
  \BibitemOpen
  \bibfield  {author} {\bibinfo {author} {\bibfnamefont {S.}~\bibnamefont
  {Nakajima}},\ }\href@noop {} {\bibfield  {journal} {\bibinfo  {journal}
  {Prog. Theor. Phys.}\ }\textbf {\bibinfo {volume} {20}},\ \bibinfo {pages}
  {948} (\bibinfo {year} {1958})}\BibitemShut {NoStop}%
\bibitem [{\citenamefont {Zwanzig}(1960)}]{Zwanzig1960}%
  \BibitemOpen
  \bibfield  {author} {\bibinfo {author} {\bibfnamefont {R.}~\bibnamefont
  {Zwanzig}},\ }\href@noop {} {\bibfield  {journal} {\bibinfo  {journal} {J.
  Chem. Phys.}\ }\textbf {\bibinfo {volume} {33}},\ \bibinfo {pages} {1338}
  (\bibinfo {year} {1960})}\BibitemShut {NoStop}%
\bibitem [{\citenamefont {Mori}(1965)}]{Mori1965}%
  \BibitemOpen
  \bibfield  {author} {\bibinfo {author} {\bibfnamefont {H.}~\bibnamefont
  {Mori}},\ }\href@noop {} {\bibfield  {journal} {\bibinfo  {journal} {Prog.
  Theor. Phys.}\ }\textbf {\bibinfo {volume} {33}},\ \bibinfo {pages} {423}
  (\bibinfo {year} {1965})}\BibitemShut {NoStop}%
\bibitem [{\citenamefont {Kidon}\ \emph {et~al.}(2018)\citenamefont {Kidon},
  \citenamefont {Wang}, \citenamefont {Thoss},\ and\ \citenamefont
  {Rabani}}]{Kidon2018}%
  \BibitemOpen
  \bibfield  {author} {\bibinfo {author} {\bibfnamefont {L.}~\bibnamefont
  {Kidon}}, \bibinfo {author} {\bibfnamefont {H.}~\bibnamefont {Wang}},
  \bibinfo {author} {\bibfnamefont {M.}~\bibnamefont {Thoss}}, \ and\ \bibinfo
  {author} {\bibfnamefont {E.}~\bibnamefont {Rabani}},\ }\href@noop {}
  {\bibfield  {journal} {\bibinfo  {journal} {J. Chem. Phys.}\ }\textbf
  {\bibinfo {volume} {149}},\ \bibinfo {pages} {104105} (\bibinfo {year}
  {2018})}\BibitemShut {NoStop}%
\bibitem [{\citenamefont {Breuer}\ \emph {et~al.}(2016)\citenamefont {Breuer},
  \citenamefont {Laine}, \citenamefont {Piilo},\ and\ \citenamefont
  {Vacchini}}]{Breuer2016}%
  \BibitemOpen
  \bibfield  {author} {\bibinfo {author} {\bibfnamefont {H.-P.}\ \bibnamefont
  {Breuer}}, \bibinfo {author} {\bibfnamefont {E.-M.}\ \bibnamefont {Laine}},
  \bibinfo {author} {\bibfnamefont {J.}~\bibnamefont {Piilo}}, \ and\ \bibinfo
  {author} {\bibfnamefont {B.}~\bibnamefont {Vacchini}},\ }\href@noop {}
  {\bibfield  {journal} {\bibinfo  {journal} {Rev. Mod. Phys.}\ }\textbf
  {\bibinfo {volume} {88}},\ \bibinfo {pages} {021002} (\bibinfo {year}
  {2016})}\BibitemShut {NoStop}%
\bibitem [{\citenamefont {Wolf}\ \emph {et~al.}(2008)\citenamefont {Wolf},
  \citenamefont {Eisert}, \citenamefont {Cubitt},\ and\ \citenamefont
  {Cirac}}]{Wolf2008}%
  \BibitemOpen
  \bibfield  {author} {\bibinfo {author} {\bibfnamefont {M.~M.}\ \bibnamefont
  {Wolf}}, \bibinfo {author} {\bibfnamefont {J.}~\bibnamefont {Eisert}},
  \bibinfo {author} {\bibfnamefont {T.~S.}\ \bibnamefont {Cubitt}}, \ and\
  \bibinfo {author} {\bibfnamefont {J.~I.}\ \bibnamefont {Cirac}},\ }\href@noop
  {} {\bibfield  {journal} {\bibinfo  {journal} {Phys. Rev. Lett.}\ }\textbf
  {\bibinfo {volume} {101}},\ \bibinfo {pages} {150402} (\bibinfo {year}
  {2008})}\BibitemShut {NoStop}%
\bibitem [{\citenamefont {Rivas}\ \emph {et~al.}(2010)\citenamefont {Rivas},
  \citenamefont {Huelga},\ and\ \citenamefont {Plenio}}]{Rivas2010}%
  \BibitemOpen
  \bibfield  {author} {\bibinfo {author} {\bibfnamefont {A.}~\bibnamefont
  {Rivas}}, \bibinfo {author} {\bibfnamefont {S.~F.}\ \bibnamefont {Huelga}}, \
  and\ \bibinfo {author} {\bibfnamefont {M.~B.}\ \bibnamefont {Plenio}},\
  }\href@noop {} {\bibfield  {journal} {\bibinfo  {journal} {Phys. Rev. Lett.}\
  }\textbf {\bibinfo {volume} {105}},\ \bibinfo {pages} {050403} (\bibinfo
  {year} {2010})}\BibitemShut {NoStop}%
\bibitem [{\citenamefont {Chru\ifmmode \acute{s}\else
  \'{s}\fi{}ci\ifmmode~\acute{n}\else \'{n}\fi{}ski}\ \emph
  {et~al.}(2011)\citenamefont {Chru\ifmmode \acute{s}\else
  \'{s}\fi{}ci\ifmmode~\acute{n}\else \'{n}\fi{}ski}, \citenamefont
  {Kossakowski},\ and\ \citenamefont {Rivas}}]{Chruscinski2011}%
  \BibitemOpen
  \bibfield  {author} {\bibinfo {author} {\bibfnamefont {D.}~\bibnamefont
  {Chru\ifmmode \acute{s}\else \'{s}\fi{}ci\ifmmode~\acute{n}\else
  \'{n}\fi{}ski}}, \bibinfo {author} {\bibfnamefont {A.}~\bibnamefont
  {Kossakowski}}, \ and\ \bibinfo {author} {\bibfnamefont {A.}~\bibnamefont
  {Rivas}},\ }\href@noop {} {\bibfield  {journal} {\bibinfo  {journal} {Phys.
  Rev. A}\ }\textbf {\bibinfo {volume} {83}},\ \bibinfo {pages} {052128}
  (\bibinfo {year} {2011})}\BibitemShut {NoStop}%
\bibitem [{\citenamefont {Luo}\ \emph {et~al.}(2012)\citenamefont {Luo},
  \citenamefont {Fu},\ and\ \citenamefont {Song}}]{Luo2012a}%
  \BibitemOpen
  \bibfield  {author} {\bibinfo {author} {\bibfnamefont {S.}~\bibnamefont
  {Luo}}, \bibinfo {author} {\bibfnamefont {S.}~\bibnamefont {Fu}}, \ and\
  \bibinfo {author} {\bibfnamefont {H.}~\bibnamefont {Song}},\ }\href@noop {}
  {\bibfield  {journal} {\bibinfo  {journal} {Phys. Rev. A}\ }\textbf {\bibinfo
  {volume} {86}},\ \bibinfo {pages} {044101} (\bibinfo {year}
  {2012})}\BibitemShut {NoStop}%
\bibitem [{\citenamefont {Breuer}\ \emph {et~al.}(2009)\citenamefont {Breuer},
  \citenamefont {Laine},\ and\ \citenamefont {Piilo}}]{Breuer2009}%
  \BibitemOpen
  \bibfield  {author} {\bibinfo {author} {\bibfnamefont {H.-P.}\ \bibnamefont
  {Breuer}}, \bibinfo {author} {\bibfnamefont {E.-M.}\ \bibnamefont {Laine}}, \
  and\ \bibinfo {author} {\bibfnamefont {J.}~\bibnamefont {Piilo}},\
  }\href@noop {} {\bibfield  {journal} {\bibinfo  {journal} {Phys. Rev. Lett.}\
  }\textbf {\bibinfo {volume} {103}},\ \bibinfo {pages} {210401} (\bibinfo
  {year} {2009})}\BibitemShut {NoStop}%
\bibitem [{\citenamefont {Laine}\ \emph {et~al.}(2010)\citenamefont {Laine},
  \citenamefont {Piilo},\ and\ \citenamefont {Breuer}}]{Laine2010a}%
  \BibitemOpen
  \bibfield  {author} {\bibinfo {author} {\bibfnamefont {E.-M.}\ \bibnamefont
  {Laine}}, \bibinfo {author} {\bibfnamefont {J.}~\bibnamefont {Piilo}}, \ and\
  \bibinfo {author} {\bibfnamefont {H.-P.}\ \bibnamefont {Breuer}},\
  }\href@noop {} {\bibfield  {journal} {\bibinfo  {journal} {Phys. Rev. A}\
  }\textbf {\bibinfo {volume} {81}},\ \bibinfo {pages} {062115} (\bibinfo
  {year} {2010})}\BibitemShut {NoStop}%
\bibitem [{\citenamefont {Nielsen}\ and\ \citenamefont
  {Chuang}(2000)}]{Nielsen2000}%
  \BibitemOpen
  \bibfield  {author} {\bibinfo {author} {\bibfnamefont {M.~A.}\ \bibnamefont
  {Nielsen}}\ and\ \bibinfo {author} {\bibfnamefont {I.~L.}\ \bibnamefont
  {Chuang}},\ }\href@noop {} {\emph {\bibinfo {title} {Quantum Computation and
  Quantum Information}}}\ (\bibinfo  {publisher} {Cambridge University Press,
  Cambridge,England},\ \bibinfo {year} {2000})\BibitemShut {NoStop}%
\bibitem [{\citenamefont {Hayashi}(2006)}]{Hayashi2006}%
  \BibitemOpen
  \bibfield  {author} {\bibinfo {author} {\bibfnamefont {M.}~\bibnamefont
  {Hayashi}},\ }\href@noop {} {\emph {\bibinfo {title} {Quantum Information}}}\
  (\bibinfo  {publisher} {Springer, Berlin},\ \bibinfo {year}
  {2006})\BibitemShut {NoStop}%
\bibitem [{\citenamefont {Fuchs}\ and\ \citenamefont {van~de
  Graaf}(1999)}]{Fuchs1999}%
  \BibitemOpen
  \bibfield  {author} {\bibinfo {author} {\bibfnamefont {C.~A.}\ \bibnamefont
  {Fuchs}}\ and\ \bibinfo {author} {\bibfnamefont {J.}~\bibnamefont {van~de
  Graaf}},\ }\href@noop {} {\bibfield  {journal} {\bibinfo  {journal} {IEEE
  Trans. Inf. Theory}\ }\textbf {\bibinfo {volume} {45}},\ \bibinfo {pages}
  {1216} (\bibinfo {year} {1999})}\BibitemShut {NoStop}%
\bibitem [{\citenamefont {Ruskai}(1994)}]{Ruskai1994}%
  \BibitemOpen
  \bibfield  {author} {\bibinfo {author} {\bibfnamefont {M.~B.}\ \bibnamefont
  {Ruskai}},\ }\href@noop {} {\bibfield  {journal} {\bibinfo  {journal} {Rev.
  Math. Phys.}\ }\textbf {\bibinfo {volume} {06}},\ \bibinfo {pages} {1147}
  (\bibinfo {year} {1994})}\BibitemShut {NoStop}%
\bibitem [{\citenamefont {Clos}\ and\ \citenamefont {Breuer}(2012)}]{Clos2012}%
  \BibitemOpen
  \bibfield  {author} {\bibinfo {author} {\bibfnamefont {G.}~\bibnamefont
  {Clos}}\ and\ \bibinfo {author} {\bibfnamefont {H.-P.}\ \bibnamefont
  {Breuer}},\ }\href@noop {} {\bibfield  {journal} {\bibinfo  {journal} {Phys.
  Rev. A}\ }\textbf {\bibinfo {volume} {86}},\ \bibinfo {pages} {012115}
  (\bibinfo {year} {2012})}\BibitemShut {NoStop}%
\bibitem [{\citenamefont {Chen}\ \emph {et~al.}(2015)\citenamefont {Chen},
  \citenamefont {Lambert}, \citenamefont {Cheng}, \citenamefont {Chen},\ and\
  \citenamefont {Nori}}]{Chen15}%
  \BibitemOpen
  \bibfield  {author} {\bibinfo {author} {\bibfnamefont {H.-B.}\ \bibnamefont
  {Chen}}, \bibinfo {author} {\bibfnamefont {N.}~\bibnamefont {Lambert}},
  \bibinfo {author} {\bibfnamefont {Y.-C.}\ \bibnamefont {Cheng}}, \bibinfo
  {author} {\bibfnamefont {Y.-N.}\ \bibnamefont {Chen}}, \ and\ \bibinfo
  {author} {\bibfnamefont {F.}~\bibnamefont {Nori}},\ }\href@noop {} {\bibfield
   {journal} {\bibinfo  {journal} {Sci. Rep.}\ }\textbf {\bibinfo {volume}
  {5}},\ \bibinfo {pages} {12753} (\bibinfo {year} {2015})}\BibitemShut
  {NoStop}%
\bibitem [{\citenamefont {Rivas}(2017)}]{Rivas17}%
  \BibitemOpen
  \bibfield  {author} {\bibinfo {author} {\bibfnamefont {A.}~\bibnamefont
  {Rivas}},\ }\href@noop {} {\bibfield  {journal} {\bibinfo  {journal} {Phys.
  Rev. A}\ }\textbf {\bibinfo {volume} {95}},\ \bibinfo {pages} {042104}
  (\bibinfo {year} {2017})}\BibitemShut {NoStop}%
\bibitem [{\citenamefont {Hinarejos}\ \emph {et~al.}(2017)\citenamefont
  {Hinarejos}, \citenamefont {Banuls}, \citenamefont {Perez},\ and\
  \citenamefont {{de Vega}}}]{Hinarejos17}%
  \BibitemOpen
  \bibfield  {author} {\bibinfo {author} {\bibfnamefont {M.}~\bibnamefont
  {Hinarejos}}, \bibinfo {author} {\bibfnamefont {M.-C.}\ \bibnamefont
  {Banuls}}, \bibinfo {author} {\bibfnamefont {A.}~\bibnamefont {Perez}}, \
  and\ \bibinfo {author} {\bibfnamefont {I.}~\bibnamefont {{de Vega}}},\
  }\href@noop {} {\bibfield  {journal} {\bibinfo  {journal} {J. Phys. A: Math.
  Theor.}\ }\textbf {\bibinfo {volume} {50}},\ \bibinfo {pages} {335301}
  (\bibinfo {year} {2017})}\BibitemShut {NoStop}%
\bibitem [{\citenamefont {Kurt}\ and\ \citenamefont {Eryigit}(2018)}]{Kurt18}%
  \BibitemOpen
  \bibfield  {author} {\bibinfo {author} {\bibfnamefont {A.}~\bibnamefont
  {Kurt}}\ and\ \bibinfo {author} {\bibfnamefont {R.}~\bibnamefont {Eryigit}},\
  }\href@noop {} {\bibfield  {journal} {\bibinfo  {journal} {Phys. Rev. A}\
  }\textbf {\bibinfo {volume} {98}},\ \bibinfo {pages} {042125} (\bibinfo
  {year} {2018})}\BibitemShut {NoStop}%
\bibitem [{\citenamefont {Mujica-Martinez}\ \emph {et~al.}(2013)\citenamefont
  {Mujica-Martinez}, \citenamefont {Nalbach},\ and\ \citenamefont
  {Thorwart}}]{Thorwart2013}%
  \BibitemOpen
  \bibfield  {author} {\bibinfo {author} {\bibfnamefont {C.~A.}\ \bibnamefont
  {Mujica-Martinez}}, \bibinfo {author} {\bibfnamefont {P.}~\bibnamefont
  {Nalbach}}, \ and\ \bibinfo {author} {\bibfnamefont {M.}~\bibnamefont
  {Thorwart}},\ }\href@noop {} {\bibfield  {journal} {\bibinfo  {journal}
  {Phys. Rev. E}\ }\textbf {\bibinfo {volume} {88}},\ \bibinfo {pages} {062719}
  (\bibinfo {year} {2013})}\BibitemShut {NoStop}%
\bibitem [{\citenamefont {Einsiedler}\ \emph {et~al.}(2020)\citenamefont
  {Einsiedler}, \citenamefont {Ketterer},\ and\ \citenamefont
  {Breuer}}]{Einsiedler20}%
  \BibitemOpen
  \bibfield  {author} {\bibinfo {author} {\bibfnamefont {S.}~\bibnamefont
  {Einsiedler}}, \bibinfo {author} {\bibfnamefont {A.}~\bibnamefont
  {Ketterer}}, \ and\ \bibinfo {author} {\bibfnamefont {H.-P.}\ \bibnamefont
  {Breuer}},\ }\href@noop {} {\bibfield  {journal} {\bibinfo  {journal} {Phys.
  Rev. A}\ }\textbf {\bibinfo {volume} {102}},\ \bibinfo {pages} {022228}
  (\bibinfo {year} {2020})}\BibitemShut {NoStop}%
\bibitem [{\citenamefont {Leggett}\ \emph {et~al.}(1987)\citenamefont
  {Leggett}, \citenamefont {Chakravarty}, \citenamefont {Dorsey}, \citenamefont
  {Fisher}, \citenamefont {Garg},\ and\ \citenamefont {Zwerger}}]{Leggett1987}%
  \BibitemOpen
  \bibfield  {author} {\bibinfo {author} {\bibfnamefont {A.~J.}\ \bibnamefont
  {Leggett}}, \bibinfo {author} {\bibfnamefont {S.}~\bibnamefont
  {Chakravarty}}, \bibinfo {author} {\bibfnamefont {A.~T.}\ \bibnamefont
  {Dorsey}}, \bibinfo {author} {\bibfnamefont {M.~P.}\ \bibnamefont {Fisher}},
  \bibinfo {author} {\bibfnamefont {A.}~\bibnamefont {Garg}}, \ and\ \bibinfo
  {author} {\bibfnamefont {W.}~\bibnamefont {Zwerger}},\ }\href@noop {}
  {\bibfield  {journal} {\bibinfo  {journal} {Rev. Mod. Phys.}\ }\textbf
  {\bibinfo {volume} {59}},\ \bibinfo {pages} {1} (\bibinfo {year}
  {1987})}\BibitemShut {NoStop}%
\bibitem [{\citenamefont {Marcus}\ and\ \citenamefont
  {Sutin}(1985)}]{Marcus1985}%
  \BibitemOpen
  \bibfield  {author} {\bibinfo {author} {\bibfnamefont {R.~A.}\ \bibnamefont
  {Marcus}}\ and\ \bibinfo {author} {\bibfnamefont {N.}~\bibnamefont {Sutin}},\
  }\href@noop {} {\bibfield  {journal} {\bibinfo  {journal} {Biochim. Biophys.
  Acta}\ }\textbf {\bibinfo {volume} {811}},\ \bibinfo {pages} {265 } (\bibinfo
  {year} {1985})}\BibitemShut {NoStop}%
\bibitem [{\citenamefont {Weiss}\ \emph {et~al.}(1987)\citenamefont {Weiss},
  \citenamefont {Grabert},\ and\ \citenamefont {Linkwitz}}]{Weiss1987}%
  \BibitemOpen
  \bibfield  {author} {\bibinfo {author} {\bibfnamefont {U.}~\bibnamefont
  {Weiss}}, \bibinfo {author} {\bibfnamefont {H.}~\bibnamefont {Grabert}}, \
  and\ \bibinfo {author} {\bibfnamefont {S.}~\bibnamefont {Linkwitz}},\
  }\href@noop {} {\bibfield  {journal} {\bibinfo  {journal} {J. Low Temp.
  Phys.}\ }\textbf {\bibinfo {volume} {68}},\ \bibinfo {pages} {213} (\bibinfo
  {year} {1987})}\BibitemShut {NoStop}%
\bibitem [{\citenamefont {Bray}\ and\ \citenamefont {Moore}(1982)}]{Bray1982}%
  \BibitemOpen
  \bibfield  {author} {\bibinfo {author} {\bibfnamefont {A.~J.}\ \bibnamefont
  {Bray}}\ and\ \bibinfo {author} {\bibfnamefont {M.~A.}\ \bibnamefont
  {Moore}},\ }\href@noop {} {\bibfield  {journal} {\bibinfo  {journal} {Phys.
  Rev. Lett.}\ }\textbf {\bibinfo {volume} {49}},\ \bibinfo {pages} {1545}
  (\bibinfo {year} {1982})}\BibitemShut {NoStop}%
\bibitem [{\citenamefont {Chakravarty}(1982)}]{Chakravarty1982}%
  \BibitemOpen
  \bibfield  {author} {\bibinfo {author} {\bibfnamefont {S.}~\bibnamefont
  {Chakravarty}},\ }\href@noop {} {\bibfield  {journal} {\bibinfo  {journal}
  {Phys. Rev. Lett.}\ }\textbf {\bibinfo {volume} {49}},\ \bibinfo {pages}
  {681} (\bibinfo {year} {1982})}\BibitemShut {NoStop}%
\bibitem [{\citenamefont {Wang}\ and\ \citenamefont {Shao}(2019)}]{Wang2019}%
  \BibitemOpen
  \bibfield  {author} {\bibinfo {author} {\bibfnamefont {H.}~\bibnamefont
  {Wang}}\ and\ \bibinfo {author} {\bibfnamefont {J.}~\bibnamefont {Shao}},\
  }\href@noop {} {\bibfield  {journal} {\bibinfo  {journal} {J. Phys. Chem. A}\
  }\textbf {\bibinfo {volume} {123}},\ \bibinfo {pages} {1882} (\bibinfo {year}
  {2019})}\BibitemShut {NoStop}%
\bibitem [{\citenamefont {Thoss}\ \emph {et~al.}(2001)\citenamefont {Thoss},
  \citenamefont {Wang},\ and\ \citenamefont {Miller}}]{Thoss2001}%
  \BibitemOpen
  \bibfield  {author} {\bibinfo {author} {\bibfnamefont {M.}~\bibnamefont
  {Thoss}}, \bibinfo {author} {\bibfnamefont {H.}~\bibnamefont {Wang}}, \ and\
  \bibinfo {author} {\bibfnamefont {W.~H.}\ \bibnamefont {Miller}},\
  }\href@noop {} {\bibfield  {journal} {\bibinfo  {journal} {J. Chem. Phys.}\
  }\textbf {\bibinfo {volume} {115}},\ \bibinfo {pages} {2991} (\bibinfo {year}
  {2001})}\BibitemShut {NoStop}%
\bibitem [{\citenamefont {Wang}\ and\ \citenamefont {Thoss}(2008)}]{Wang2008}%
  \BibitemOpen
  \bibfield  {author} {\bibinfo {author} {\bibfnamefont {H.}~\bibnamefont
  {Wang}}\ and\ \bibinfo {author} {\bibfnamefont {M.}~\bibnamefont {Thoss}},\
  }\href@noop {} {\bibfield  {journal} {\bibinfo  {journal} {New J. Phys.}\
  }\textbf {\bibinfo {volume} {10}},\ \bibinfo {pages} {115005} (\bibinfo
  {year} {2008})}\BibitemShut {NoStop}%
\bibitem [{\citenamefont {Wang}\ and\ \citenamefont {Thoss}(2010)}]{Wang10}%
  \BibitemOpen
  \bibfield  {author} {\bibinfo {author} {\bibfnamefont {H.}~\bibnamefont
  {Wang}}\ and\ \bibinfo {author} {\bibfnamefont {M.}~\bibnamefont {Thoss}},\
  }\href@noop {} {\bibfield  {journal} {\bibinfo  {journal} {Chem. Phys.}\
  }\textbf {\bibinfo {volume} {370}},\ \bibinfo {pages} {78} (\bibinfo {year}
  {2010})}\BibitemShut {NoStop}%
\bibitem [{\citenamefont {Wang}\ and\ \citenamefont {Thoss}(2003)}]{Wang2003}%
  \BibitemOpen
  \bibfield  {author} {\bibinfo {author} {\bibfnamefont {H.}~\bibnamefont
  {Wang}}\ and\ \bibinfo {author} {\bibfnamefont {M.}~\bibnamefont {Thoss}},\
  }\href@noop {} {\bibfield  {journal} {\bibinfo  {journal} {J. Chem. Phys.}\
  }\textbf {\bibinfo {volume} {119}},\ \bibinfo {pages} {1289} (\bibinfo {year}
  {2003})}\BibitemShut {NoStop}%
\bibitem [{\citenamefont {Manthe}(2008)}]{Manthe08}%
  \BibitemOpen
  \bibfield  {author} {\bibinfo {author} {\bibfnamefont {U.}~\bibnamefont
  {Manthe}},\ }\href@noop {} {\bibfield  {journal} {\bibinfo  {journal} {J.
  Chem. Phys.}\ }\textbf {\bibinfo {volume} {128}},\ \bibinfo {pages} {164116}
  (\bibinfo {year} {2008})}\BibitemShut {NoStop}%
\bibitem [{\citenamefont {Vendrell}\ and\ \citenamefont
  {Meyer}(2011)}]{Vendrell11}%
  \BibitemOpen
  \bibfield  {author} {\bibinfo {author} {\bibfnamefont {O.}~\bibnamefont
  {Vendrell}}\ and\ \bibinfo {author} {\bibfnamefont {H.-D.}\ \bibnamefont
  {Meyer}},\ }\href@noop {} {\bibfield  {journal} {\bibinfo  {journal} {J.
  Chem. Phys.}\ }\textbf {\bibinfo {volume} {134}},\ \bibinfo {pages} {044135}
  (\bibinfo {year} {2011})}\BibitemShut {NoStop}%
\bibitem [{\citenamefont {Wang}(2015)}]{Wang2015}%
  \BibitemOpen
  \bibfield  {author} {\bibinfo {author} {\bibfnamefont {H.}~\bibnamefont
  {Wang}},\ }\href@noop {} {\bibfield  {journal} {\bibinfo  {journal} {J. Phys.
  Chem. A}\ }\textbf {\bibinfo {volume} {119}},\ \bibinfo {pages} {7951}
  (\bibinfo {year} {2015})}\BibitemShut {NoStop}%
\bibitem [{\citenamefont {Frenkel}(1934)}]{Frenke1934}%
  \BibitemOpen
  \bibfield  {author} {\bibinfo {author} {\bibfnamefont {J.}~\bibnamefont
  {Frenkel}},\ }\href@noop {} {\emph {\bibinfo {title} {Wave Mechanics}}}\
  (\bibinfo  {publisher} {Oxford: Clarendon},\ \bibinfo {year}
  {1934})\BibitemShut {NoStop}%
\bibitem [{\citenamefont {Wang}\ and\ \citenamefont {Thoss}(2009)}]{Wang2009}%
  \BibitemOpen
  \bibfield  {author} {\bibinfo {author} {\bibfnamefont {H.}~\bibnamefont
  {Wang}}\ and\ \bibinfo {author} {\bibfnamefont {M.}~\bibnamefont {Thoss}},\
  }\href@noop {} {\bibfield  {journal} {\bibinfo  {journal} {J. Chem. Phys.}\
  }\textbf {\bibinfo {volume} {131}},\ \bibinfo {pages} {024114} (\bibinfo
  {year} {2009})}\BibitemShut {NoStop}%
\bibitem [{\citenamefont {de~Vega}\ \emph {et~al.}(2015)\citenamefont
  {de~Vega}, \citenamefont {Schollw\"ock},\ and\ \citenamefont
  {Wolf}}]{Vega2015}%
  \BibitemOpen
  \bibfield  {author} {\bibinfo {author} {\bibfnamefont {I.}~\bibnamefont
  {de~Vega}}, \bibinfo {author} {\bibfnamefont {U.}~\bibnamefont
  {Schollw\"ock}}, \ and\ \bibinfo {author} {\bibfnamefont {F.~A.}\
  \bibnamefont {Wolf}},\ }\href@noop {} {\bibfield  {journal} {\bibinfo
  {journal} {Phys. Rev. B}\ }\textbf {\bibinfo {volume} {92}},\ \bibinfo
  {pages} {155126} (\bibinfo {year} {2015})}\BibitemShut {NoStop}%
\bibitem [{\citenamefont {Wi\ss{}mann}\ \emph {et~al.}(2012)\citenamefont
  {Wi\ss{}mann}, \citenamefont {Karlsson}, \citenamefont {Laine}, \citenamefont
  {Piilo},\ and\ \citenamefont {Breuer}}]{Wissmann2012}%
  \BibitemOpen
  \bibfield  {author} {\bibinfo {author} {\bibfnamefont {S.}~\bibnamefont
  {Wi\ss{}mann}}, \bibinfo {author} {\bibfnamefont {A.}~\bibnamefont
  {Karlsson}}, \bibinfo {author} {\bibfnamefont {E.-M.}\ \bibnamefont {Laine}},
  \bibinfo {author} {\bibfnamefont {J.}~\bibnamefont {Piilo}}, \ and\ \bibinfo
  {author} {\bibfnamefont {H.-P.}\ \bibnamefont {Breuer}},\ }\href@noop {}
  {\bibfield  {journal} {\bibinfo  {journal} {Phys. Rev. A}\ }\textbf {\bibinfo
  {volume} {86}},\ \bibinfo {pages} {062108} (\bibinfo {year}
  {2012})}\BibitemShut {NoStop}%
\end{thebibliography}
\end{document}


\begin{center}
{\bf Supplemental Material for ``Non-Markovian effects in the spin-boson model at zero temperature''}\\
\smallskip
{S. Wenderoth,$^{1}$ H.-P. Breuer,$^{1, 2}$ M.\ Thoss$^{1, 2}$} \\
\vspace*{0.5em}
\textit{\footnotesize  $^{1}$ Institute of Physics, University of Freiburg, Hermann-Herder-Str.\ 3, D-79104 Freiburg, Germany}\\
\textit{\footnotesize $^{2}$ EUCOR Centre for Quantum Science and Quantum Computing, University of Freiburg, Hermann-Herder-Str. 3, D-79104 Freiburg, Germany}
\end{center}

\section{Mapping of the time-evolution of the two different initial states}
The time-dependent trace distance for any pair of initial states of the spin can be calculated using 
\begin{align}
\mathcal{D}(t) &= \frac{1}{2}\sqrt{ [\braket{\sigma_x}_1(t)-\braket{\sigma_x}_2(t)]^2 + [\braket{\sigma_y}_1(t)-\braket{\sigma_y}_2(t)]^2 + [\braket{\sigma_z}_1(t)-\braket{\sigma_z}_2(t)]^2},\label{eq:tdp_trace_distance}
\end{align}
where $\sigma_{x,y,z}$ denote the three Pauli matrices, and the index refers to the first and second initial state of the spin $\rho_1$ and $\rho_2$, respectively. The time-dependent expectation values in Eq.\,(\ref{eq:tdp_trace_distance}) can be related to each other, which we will employ in the following. For the special choice of $\rho_1=\ket{\uparrow}\bra{\uparrow}$ and $\rho_2=\ket{\downarrow}\bra{\downarrow}$, in particular, the expectation values $\braket{\sigma_i}_1$ can be directly related to the expectation values $\braket{\sigma_i}_2$. In order to show this, we assume that the initial states of the joint system factorizes between the spin and the environment, and thus, can be written as
\begin{align}
\varrho_1(0) &= \ket{\uparrow}\bra{\uparrow} \otimes \rho_B(0) \\
\varrho_2(0) &= \ket{\downarrow}\bra{\downarrow} \otimes \rho_B(0),
\end{align}
where $\rho_B(0)$ is the initial state of the environment. Formally, the time-dependent expectation values of the spin are defined as
\begin{align}
\braket{\sigma_i}_{n} &= {\rm tr} \{ \sigma_i {\rm e}^{-iHt} \varrho_n(0) {\rm e}^{iHt} \},\label{eq:tdp_exp_val}
\end{align}
where $n\in \{1,2\}$, and $H$ denotes the Hamiltonian of the full system given by
\begin{align}
H &= \Delta \sigma_x + H_B + \sigma_z \sum_n c_n q_n.\label{eq:full_Hamiltonian}
\end{align}
Since $\sigma_x^2=\mathbf{1}$ and $\sigma_x^\dagger = \sigma_x$ holds, the transformation $\tilde{O}= \sigma_x O \sigma_x$ is unitary. Expectation values are invariant under unitary transformations, and thus, Eq.\,(\ref{eq:tdp_exp_val}) can be transformed to 
\begin{align}
\braket{\sigma_i}_{1/2} &= {\rm tr} \{ \sigma_x \sigma_i\sigma_x \sigma_x {\rm e}^{-iHt}\sigma_x  \sigma_x \rho_{1/2}(0)\sigma_x \sigma_x {\rm e}^{iHt}\sigma_x \}\\
 &= {\rm tr} \{ \tilde{\sigma_i}{\rm e}^{-i\tilde{H}t} \tilde{\rho}_{1/2}(0){\rm e}^{i\tilde{H}t} \} \nonumber.
\end{align}
This transformation only acts on the Hilbert space of the spin, and thus, transforms only the spin operators. The transformed operators and initial state read
\begin{align}
\sigma_x^\dagger \sigma_x \sigma_x = \sigma_x, \\
\sigma_x^\dagger \sigma_y \sigma_x = -\sigma_y, \\
\sigma_x^\dagger \sigma_z \sigma_x = -\sigma_z, \\
\sigma_x \rho_1(0)\sigma_x = \rho_2(0). \label{eq:transformed_initial_state}
\end{align}
Plugging this into Eq.\,(\ref{eq:full_Hamiltonian}) yields the transformed Hamiltonian
\begin{align}
\tilde{H} = \sigma_x^\dagger H \sigma_x = \Delta \sigma_x + H_B + \sigma_z \sum_n (-c_n) q_n.
\end{align}
The transformation only changes the sign of the couplings $c_n$. The properties of the environment which influences the dynamics of the spin are fully characterized by the spectral density\cite{Weiss1999,Leggett1987}, which is defined as
\begin{align}
J(\omega) &= \frac{\pi}{2} \sum_n \frac{c_n^2}{\omega_n} \delta(\omega - \omega_n).
\end{align}
Since the spectral density depends on the squared couplings, the two Hamiltonians $H$ and $\tilde{H}$ give rise to the same spectral density. Consequently, the reduced spin dynamics induced by $H$ and $\tilde{H}$ are equal. Using this and Eq.\,(\ref{eq:transformed_initial_state}) we conclude that
\begin{align}
\braket{\sigma_x}_{2}(t) &= \braket{\sigma_x}_{1}(t) \\
\braket{\sigma_y}_{2}(t) &= -\braket{\sigma_y}_{1}(t) \\
\braket{\sigma_z}_{2}(t) &= -\braket{\sigma_z}_{1}(t).
\end{align}
Employing this result, Eq.\,(\ref{eq:tdp_trace_distance}) can be expressed as
\begin{align}
\mathcal{D}(t) &= \sqrt{ \braket{\sigma_z}_1^2 + \braket{\sigma_y}_1^2 }.
\end{align}
Here and in the following we suppress the time dependence of the expectation values. Finally, we employ that for the unbiased spin-boson model $\braket{\sigma_y}= \frac{1}{2\Delta} \partial_t\braket{\sigma_z}$ holds. With this we arrive at the final equation for the trace distance
\begin{align}
\mathcal{D}(t) &= \sqrt{ \braket{\sigma_z}_1^2 +\frac{1}{(2\Delta)^2} \big[\partial_t \braket{\sigma_z}_1\big]^2 },\label{eq:trace_distance}
\end{align}
from which one can calculate the derivative of the trace distance as
\begin{align}
\partial_t \mathcal{D}(t) &=\frac{ [\partial_t \braket{\sigma_z}_1] \big[ \braket{\sigma_z}_1 + \frac{1}{(2\Delta)^2}\partial^2_t\braket{\sigma_z}_1 \big] }{\mathcal{D}(t)}. \label{eq:trace_distance_derivative}
\end{align}
From this equation it directly follows that $\partial_t\braket{\sigma_z}=0$ implies $\partial_t\mathcal{D}=0$.
\section{Analytic expression for the weak coupling regime}
In the weak coupling and large $\omega_c$ limit, an approximate analytic solution\cite{Weiss1999} for $\braket{\sigma_z}(t)$ can be derived using a path integral approach. The time evolution of $\braket{\sigma_z}$ within this approach is given by
\begin{align}
\braket{\sigma_z}(t) &= {\rm e}^{-\gamma t} \big[ \cos(\tilde{\Delta}t) + \frac{\gamma}{\tilde{\Delta}} \sin(\tilde{\Delta}t) \big] ,\label{eq:sigma_z_weak_coupling}
\end{align}
where the renormalized frequency $\tilde{\Delta}$ and the damping $\gamma$ depend on the characteristic bath frequency $\omega_c$ and the coupling strength $\alpha$ and are given by\cite{Weiss1999}
\begin{align}
\tilde{\Delta} &= \big[\Gamma(1-2\alpha) \cos(\pi \alpha) \big]^{\nicefrac{1}{2(1-\alpha)}} \bigg(\frac{2 \Delta}{\omega_c} \bigg)^{\nicefrac{\alpha}{1-\alpha}} 2\Delta\\
\gamma &= \frac{\pi}{2}\alpha\tilde{\Delta}{\rm e}^{-\nicefrac{\tilde{\Delta}}{\omega_c}},
\end{align}
where $J(\omega)$ is the spectral density of the harmonic oscillators. Using this result and Eq.\,(\ref{eq:trace_distance}) an equation for the trace distance can be derived as
\begin{align}
\mathcal{D}(t) &= {\rm e}^{-\gamma t} \sqrt{\frac{1}{2}[1+\eta] + \beta \sin(2\tilde{\Delta}t) + \frac{1}{2} [1-\eta] \cos(2\tilde{\Delta}t)},\label{eq:trace_distance_weak_coupling}	
\end{align}
where we defined the following constants
\begin{align}
\beta &= \frac{\gamma}{\tilde{\Delta}},\\
\eta &= \beta^2 + \frac{\tilde{\Delta}}{2\Delta}(1+\beta)^2.
\end{align}
\section{Non-analyticity of the non-Markovianity}
Equation (\ref{eq:trace_distance_weak_coupling}) can be used to further analyse the behaviour of the non-Markovianity $\mathcal{N}$ in the weak coupling limit. First note that for $\alpha=0$ the time evolution is unitary, and thus, $\mathcal{N}=0$. To derive the behaviour in the limit $\alpha \to 0$, we note that Eq.\,(\ref{eq:trace_distance_weak_coupling}) obeys
\begin{align}
\mathcal{D}\big(t+\frac{\pi}{\tilde{\Delta}}\big) &= {\rm e}^{\frac{-\pi \gamma}{\tilde{\Delta}}}\mathcal{D}(t). \label{eq:periodicity_property}
\end{align}
Since this holds for all times $t$, the same holds for the derivative of the trace distance $\sigma(t)=\partial_t \mathcal{D}(t)$. Assuming that the non-Markovianity $\mathcal{N}$ is finite, which is true for all parameters considered here, the defining integral of $\mathcal{N}$ can be partitioned as
\begin{align}
\mathcal{N} &= \sum_{n=0}^\infty \int\displaylimits_{\substack{0 \\ \sigma>0}}^{\nicefrac{\pi}{\tilde{\Delta}}}{\rm d}t~ \sigma\big(n\frac{\pi}{\tilde{\Delta}} + t\big).\label{eq:partitioned_integral}
\end{align}
Employing property (\ref{eq:periodicity_property}) this can be written as
\begin{align}
\mathcal{N}&=\sum_{n=0}^\infty {\rm e}^{-n \pi  \nicefrac{\gamma}{\tilde{\Delta}}} \int\displaylimits_{\substack{0 \\ \sigma>0}}^{\nicefrac{\pi}{\tilde{\Delta}}} {\rm d}t~ \sigma(t),
\end{align}
where the integral measures the information back flow during the first period of the time evolution. Let $I \subset [t, t+\nicefrac{\pi}{\tilde{\Delta}}]$ denote the times at which $\sigma(t)>0$. For simplicity, we assume that $I$ consists of a single, connected interval. This is true for the analytic solution given by Eq.\,(\ref{eq:trace_distance_weak_coupling}). In the following we denote with $t_{\rm min}$ and $t_{\rm max}$ the lower and upper end of the interval $I$, respectively.  From the fundamental theorem of calculus it then follows that
\begin{align}
\mathcal{N} &= \sum_{n=0}^\infty {\rm e}^{-n \pi  \nicefrac{\gamma}{\tilde{\Delta}}} \big(\mathcal{D}(t_{max})-\mathcal{D}(t_{min}) \big).
\end{align}
Since $\pi  \nicefrac{\gamma}{\tilde{\Delta}}>0$ the geometric series can be resumed resulting in
\begin{align}
\mathcal{N} &= \frac{\mathcal{D}(t_{max})-\mathcal{D}(t_{min}) }{1-{\rm e}^{-\pi  \nicefrac{\gamma}{\tilde{\Delta}}}}. \label{eq:resumed_non-Markovianity}
\end{align}
The nominator in Eq.\,(\ref{eq:resumed_non-Markovianity}) accounts for memory effects occurring in the first period, whereas the denominator accounts for the remaining, periodically occurring, information back flows. The limit of the non-Markovianity as $\alpha\to 0$ can be obtained by considering the leading order behaviour of the nominator and denominator, which read
\begin{align}
\mathcal{D}(t_{max})-\mathcal{D}(t_{min})  &\sim - \alpha \big[\ln(\nicefrac{2\Delta}{\omega_c}) + \gamma_{EM} + \frac{\pi^2}{4} {\rm e}^{-\nicefrac{2\Delta}{\omega_c}} \big] \\
1-{\rm e}^{-\pi  \nicefrac{\gamma}{\tilde{\Delta}}} &\sim \alpha \frac{\pi^2}{2}{\rm e}^{-\nicefrac{2\Delta}{\omega_c}}.
\end{align}
Since both terms are linear in $\alpha$ as $\alpha \to 0$ we conclude that
\begin{align}
\mathcal{N} &\sim -\frac{2}{\pi^2}{\rm e}^{\nicefrac{2\Delta}{\omega_c}}\big[\ln(\nicefrac{2\Delta}{\omega_c}) + \gamma_{EMC} + \frac{\pi^2}{4} {\rm e}^{-\nicefrac{2\Delta}{\omega_c}} \big],
\end{align}
where $\gamma_{EMC}$ denotes the Euler–Mascheroni constant. This shows that $\lim_{\alpha\to0^+}\mathcal{N}>0$, and thus, $\mathcal{N}$ is not analytic at $\alpha=0$.
\section{Perturbative treatment of the dynamics}
\begin{figure}[h]
\hspace*{-0.5cm}
\includegraphics[scale=0.5]{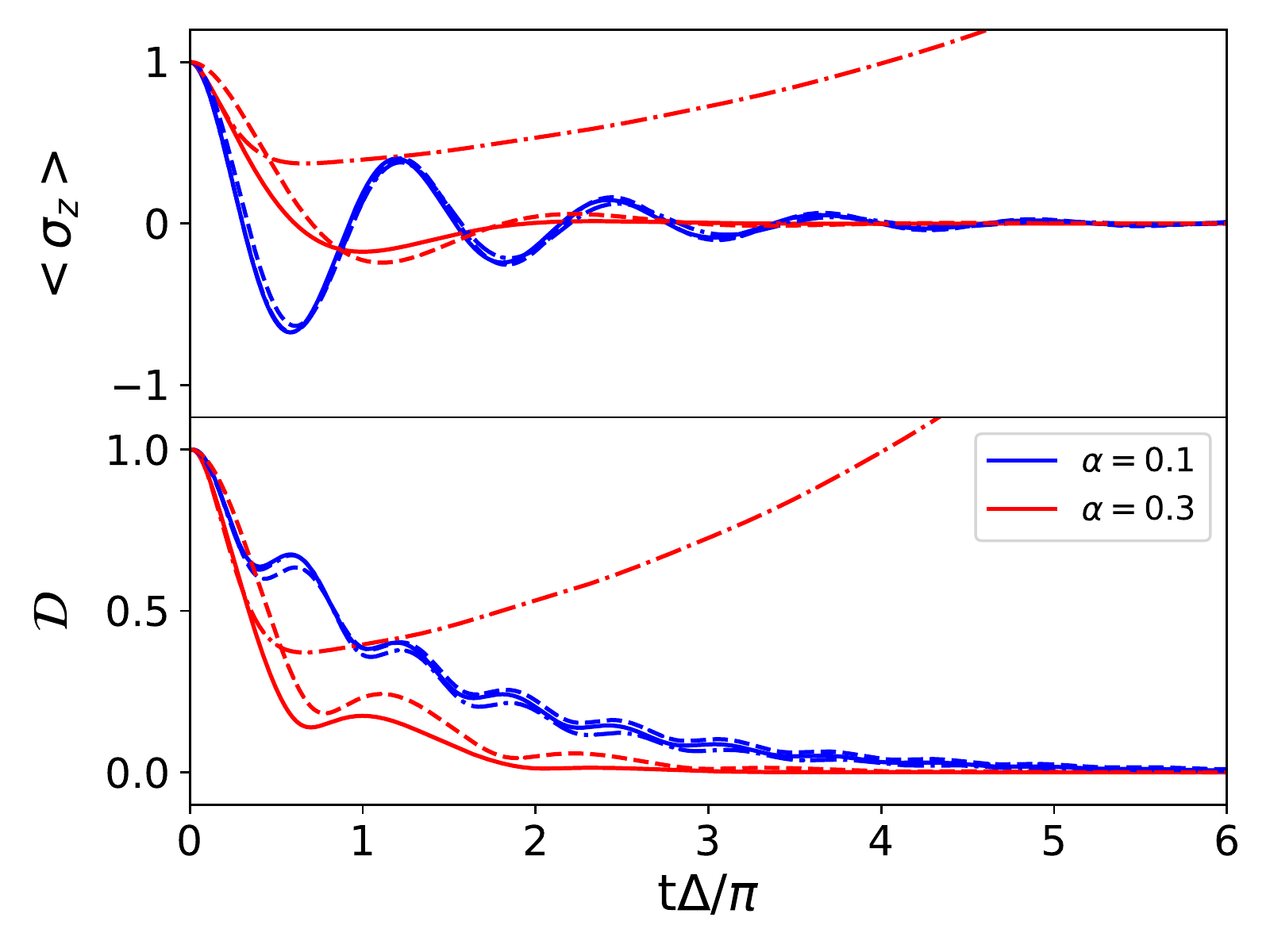}
\caption{Comparison between the numerically exact results (solid lines), the TCL2 master equation approach (dashed-dotted lines), and the analytical Eq.\,(\ref{eq:trace_distance_weak_coupling})(dashed lines), for $\alpha=0.1$ (blue curves) and $\alpha=0.3$ (red curves) for $\omega_c=20\Delta$.}\label{fig:perturbation_dynamics}
\end{figure}

Our numerically exact results allow for the validation of perturbative approaches and demonstrate certain limitations. Here, we compare the numerically exact results with the analytic solution given by Eq.\,(\ref{eq:sigma_z_weak_coupling}) and (\ref{eq:trace_distance_weak_coupling}) and results obtained with the time-convolutionless (TCL2) master equation approach.\cite{Shibata1977, Chaturvedi1979, Shibata1980} The latter was previously used in Ref.\,\onlinecite{Clos2012} to investigate memory effects in the spin-boson model. In Fig.\,\ref{fig:perturbation_dynamics}, the comparison of $\braket{\sigma_z}$ as well as the trace distance is shown for $\alpha=0.1$ and $\alpha=0.3$ for $\omega_c=20\Delta$.
For $\alpha=0.1$ the TCL2 master equation approach, as well as the analytical equation give qualitatively correct results for $\braket{\sigma_z}$ and $\mathcal{D}$. For the non-Markovianity we find $\mathcal{N} \approx 0.094$ from the TCL2 master equation approach and $\mathcal{N}\approx0.095$ from the analytic solution. Both are close to the numerically exact value of $\mathcal{N}\approx 0.088$. Note that for $\alpha=0.1$ we find this small overestimation of $\mathcal{N}$ within the two approximative approaches for all investigated values of $\omega_c>10 \Delta$. As the coupling strength increases the approximations become less accurate and eventually break down. This is exemplified for $\alpha=0.3$ in Fig.\,(\ref{fig:perturbation_dynamics}). The analytical solution still gives qualitatively good results, although the deviations from the numerically exact solution increases. For the non-Markovianity we find $\mathcal{N}\approx0.041$ from the numerically exact simulations and $\mathcal{N}\approx0.079$ from the analytical equations. The TCL2 master equation approach, on the other hand, gives non-physical results, i.e. $\braket{\sigma_z}$ and $\mathcal{D}$ exceed one after some time, demonstrating the break-down of perturbation theory.